\documentclass[onecolumn, amssymb, preprint, showpacs]{revtex4-1}   

\usepackage{lineno}

\usepackage{graphicx}
\usepackage{dcolumn}
\usepackage{amsmath,amsfonts}
\usepackage{tipa}%
\usepackage{bm}
\usepackage{subfigure}
\begin{document}
\begin{center}
\textbf { \large {Intrinsic normalization and extrinsic denormalization of formant data of vowels}}

T V Ananthapadmanabha
	\\tva.blr@gmail.com
	\\Voice and Speech Systems, Malleswaram, Bangalore, India
	\\and
	\\A G Ramakrishnan
	\\ramkiag@ee.iisc.ernet.in
	\\Indian Institute of Science, Bangalore, India%
\end{center}





\date{\today}


\textbf{Abstract:} Using a known speaker-intrinsic normalization procedure, formant data are scaled by the reciprocal of the geometric mean of the first three formant frequencies. This reduces the influence of the talker but results in a distorted vowel space. The proposed speaker-extrinsic procedure re-scales the normalized values by the mean formant values of vowels. When tested on the formant data of vowels published by Peterson and Barney, the combined approach leads to well separated clusters by reducing the spread due to talkers. The proposed procedure performs better than two top-ranked normalization procedures based on the accuracy of vowel classification as the objective measure. 



\section{INTRODUCTION}
Formant frequencies measured over the mid-part of a vowel of American English, spoken in the same context (/hVd/) by talkers of different age and gender, unanimously labelled by native listeners, show a considerable spread in the $F_2$ \textit{versus} $F_1$ space \cite{PnB}. This has motivated researchers to look for a suitable transformation or normalization of the measured  raw formant data so as to bring out the underlying invariance of vowels.  The normalization is expected to reduce the spread in the formant data arising due to the influence of talker's gender and age, while preserving the relative mean positions of the vowels as in the original formant space \cite{Disner1980, Watt}. 

There is a huge amount of literature on vowel normalization, spanning over six decades, inhibiting a critical review in this short paper. We cite some secondary sources. Adank \cite{Adank2003} gives a review of the literature till 2003. The effectiveness of some select vowel normalization methods have been compared based on certain objective criteria \cite{Fabri, FlynnICPhs, Adank2004}. Carpenter and Govindarajan\cite{CarpandGovind} give a brief description as well as an evaluation of 32 intrinsic and 128 extrinsic procedures for the vowel classification task.  Normalization in the context of sociolinguistics has also been reported \cite{Adank2004, Fabricius2007}.   

Some important milestones of research in this area are briefly covered. 
On an average, the vocal tract length (VTL) of an adult female (or child) is shorter than that of an adult male. Theoretically, this implies that all the formants be scaled inversely as the ratio of VTLs. However, the ratio of the mean formant frequency of adult female speakers to that of male speakers is both vowel and formant dependent \cite{Fant1973}, varying over a wide range of 1.03 to 1.30 for the data  published by Peterson and Barney\cite{PnB} (abbreviated as P\&B).  This wide range  combined with the fact that the mean formant frequency of adult female speakers has been reported to be lower than that of adult male speakers for a specific Swedish vowel \cite{Fant1973}, has led researchers to speculate that factors other than VTL, such as possible gender based differences in articulation, may also contribute to the noted differences in the formant ratios\cite{KJ}. F0 also has been considered to be an additional parameter for disambiguating vowels. For normalization, researchers have proposed differences such as $(F_1-F0)$, $(F_2-F_1)$, $(F_3-F_2)$ and the ratios $(F_2/F_1)$, $(F_3/F_2)$ etc. in various frequency scales such as Koenig, log, mel or Bark \cite{SyrdalandGopal, Miller1989}. 

The topmost performing normalization procedure for automatic vowel classification yields only about 80\% accuracy even with the controlled context of P\&B data\cite{CarpandGovind}. Despite the availability of a large number of procedures, a fully satisfactory solution for normalization is yet to emerge\cite{FlynnICPhs}. This has motivated us to propose an intrinsic-cum-extrinsic normalization procedure, resulting in what we refer to as de-normalized formants. 
The effectiveness of the combined procedure in reducing the influence of talker's age and gender is illustrated using the P\&B data. Vowel classification using the pooled de-normalized formant values of all speakers (adult male, adult female and child) is shown to give a very high accuracy (95\%).  The performance of the proposed procedure compares well with, or is better than, two top-ranked normalization procedures \cite{Adank2003, Fabri}.

\section{ PROPOSED METHOD}
\subsection{Intrinsic Normalization}
The geometric mean of the first three formant frequencies \cite{Milleretal, Sussman} of a speaker's vowel sample is given by
\begin{equation}	
GM123 = [F(1)F(2)F(3)]^{(1/3)}
\end{equation}	
where $F(i)$ corresponds to the $i^{th}$ raw formant frequency in Hz. Let $AM(i)$ and $AF(i)$, $i=1,2,3$ denote the mean first three formant frequencies of adult males and females respectively. Assuming $AF(i)=\alpha AM(i)$, the ratio of geometric means, GM123(Female)/GM123(Male) is equal to $\alpha$. Hence GM123 may be expected to normalize any uniform scaling of the formant frequencies arising due to gender and age. The normalized formant frequency \cite{Milleretal, Sussman} of a given vowel sample is given by the ratio 
\begin{equation}
NF(i) =  F(i)/GM123
\end{equation}	
$NF(i)$, being a ratio, is a dimensionless quantity. Equation 2 makes use of speaker-specific data of the first three formants of only the given vowel sample. Hence the procedure has to be strictly called \textquoteleft speaker-intrinsic, formant-extrinsic, vowel-intrinsic' normalization \cite{Fabri}. Instead, for the sake of brevity, we refer to the procedure as intrinsic normalization. 

GM123 has a wide range of about 644 Hz (vowel /u/ of an adult male speaker) to 1400 Hz (vowel /\ae/ of the same speaker) for the P\&B data, i.e., a factor of more than 2. However, for a given speaker, VTL varies only by about 10\% for different vowels. The over-correction in intra-speaker normalization results in a distortion of the vowel space (See Appendix-A for illustration). Due to the very low value of GM123 for back rounded vowels, in the $NF_2$ \textit{versus} $NF_1$ space these vowels lie above vowel /\textipa{A}/ along /\textipa{A}/-/i/ direction instead of  lying below /\textipa{A}/ in /\textipa{A}/-/u/ direction as in the raw formant space. In order to restore the original relative vowel positions,  we propose an extrinsic de-normalization procedure.

\subsection{Proposed Extrinsic De-normalization Procedure}
\textbf {Assumptions:} In a normalization procedure, it is incorrect to assume the vowel identity of a sample to be known. It is for this reason that statistics of the formant data across all vowels, instead of vowel specific statistics, are generally used in the existing extrinsic procedures\cite{Adank2003, Fabri, FlynnICPhs}, \cite{KJ}. However, in this work, we make use of vowel specific statistics, the mean, $\mu(i,j)$, and the standard deviation, $\sigma(i,j)$ of vowel $j$. During the process of the proposed extrinsic normalization, the identity of the vowel sample is also determined. 

Since $\mu(i,j)$ and  $\sigma(i,j)$ depend solely on a specific formant $i$ of a specific vowel $j$, the procedure is \textquoteleft formant-intrinsic' and \textquoteleft vowel-intrinsic'\cite{Fabri}.  Since the statistics represent the average across speakers, it is \textquoteleft speaker-extrinsic'. For the sake of brevity, we use the term \textquoteleft extrinsic'.

\textbf {Development of the proposed procedure:} We define the geometric mean of the average formant frequencies for a given vowel as
\begin{equation}	
GMA123 = [\mu(1)\mu(2)\mu(3)]^{(1/3)}
\end{equation}	
(See Appendix-B for a clarification of the average formant frequency $\mu(i)$ used in the RHS of the above equation.) Initially, we explored using the ratio GMA123/GM123 as the normalization factor in Eq.(2) instead of the reciprocal of GM123. The rationale is that while GM123 is expected to normalize for the inter-speaker differences, the factor GMA123 would restore the relative vowel positions (See Appendix-B for illustration). Further, the normalized values will now have the unit of Hz, with the range of values comparable to those of the raw formant data. However, both GM123 and GMA123 are common scale factors for all the three formants of a given vowel $j$. However, as noted in Sec.I, formant ratios are both formant and vowel dependent. Hence we propose $\mu(i,j)$ itself as a scaling factor since it is both formant ($i$) and vowel ($j$) dependent. 

\textbf {Proposed extrinsic procedure:}  The intrinsic normalized formant values $NF(i)$ of a given vowel sample are transformed to what we refer to as the  de-normalized formant values. Since the vowel identity of a test sample is unknown, we use a \textquoteleft hypothesize-test' paradigm. Let $V$ be the number of vowels in the database. We hypothesize the index ($J$), one at a time, of the unknown vowel and for each hypothesis ($J$), determine the de-normalized formant values given by  	
\begin{equation}
DF(i,J) = NF(i) * \mu(i,J)
\end{equation}
In our study we find that the mapping from the dimensionless NF to DF with the unit in Hz does not affect the results \cite{Tom&Kned}. Each vowel sample $NF(i)$ maps to $V$ de-normalized values, $DF(i,J)$, for hypotheses $J=1,V$ of which only one hypothesis has to be selected. We test each hypothesis by computing the distance between the de-normalized first two formants and the mean values of the corresponding de-normalized formant data of the hypothesized vowel as
\begin{equation}
D(J)= Distance<DF(i,J),\bar\mu(i,J),\bar\sigma(i,J)>,i=1,2
\end{equation}
where Distance $<>$ denotes an appropriate distance measure. (See Sec.III.B for the definition of distance measure used and Appendix-C for a clarification of the \lq hypothesize-test' procedure and illustration of distance computation.) The third formant frequency has an indirect influence via $NF(i)$. Let $\bar{J}$ be the index for which $D(J)$ is the minimum. The vowel index is postulated as $\bar{J}$. Only  $DF(i,\bar{J})$ is taken as the de-normalized value. That is, $NF(i)$ maps to $DF(i,\bar{J})$ in the de-normalized space. This procedure at once achieves vowel de-normalization as well as vowel classification.

\textbf {A parallel to perceptual studies:} Utilizing the mean and standard deviation values implies having \textit{a priori} knowledge of the vowel space of a given language. The performance is known to degrade if anomalous information is given about the speaker's gender (male/female)\cite{JohnsonStrand1999} or the language (American English/Canadian English)\cite{Niedz1997}.  This suggests that a listener's performance of perceptual identification of vowels improves with \textit{a priori} knowledge (or familiarity) of the talker's identity or gender or language. It is speculated that listeners use a \lq cognitive frame of reference' of the talker\cite{KJ}. With this background, the use of \textit{a priori} knowledge of the mean and standard deviation values of vowel formant data appears justified. 

\subsection{ Experimental Results and Discussion}
We have used the P\&B data\cite{URL, Waltrus} for illustrating the procedure. There are 66, 56 and 30 samples for \textquoteleft men', \textquoteleft women' and \textquoteleft children' categories, respectively. We have considered all the (nine) vowels excluding the retroflex vowel /\textrhookrevepsilon/. In the illustrations to follow, a vowel triangle\cite{Fabri}$^,$\cite{FlynnICPhs}$^,$\cite{WattFrib} based on the mean values of the three corner vowels is also shown for the adult male and female speakers. Its relevance is discussed in Sec.III.A.  We have followed the convention used by P\&B in selecting the orientation of the plot with vowel /u/ near the bottom-left of the graph. In all the figures the same notation as given in Fig.1 is followed.

A plot of raw formant data,  $F_2$ \textit{versus} $F_1$ is shown in Fig.1. For the front vowels, the data show a wide spread across gender and age. Also, a considerable spread is seen within each vowel. The front vowels are not well separated and some back vowels (/\textipa{U}/ and /\textipa{u}/, /\textipa{A}/ and /\textipa{O}/) heavily overlap. Also see Fig.8 of Peterson and Barney\cite{PnB}; Fig.3 of Miller, 1989\cite{Miller1989}.

In the de-normalized formant space both the inter and intra speaker spread is reduced considerably ($DF_2$ \textit{versus} $DF_1$ plot of Fig. 2). The relative positions of vowels are preserved as in the raw formant data space. Tense/lax and high/low front vowels form distinct clusters. The separation amongst back vowels is surprisingly good. Clusters for vowels (/\textipa{U}/ and /\textipa{u}/) and (/\textipa{A}/ and /\textipa{O}/) are also reasonably well separated. 
\begin{figure}
	\centering	
	 \includegraphics[width=1\textwidth, height=.5\textheight]
	 {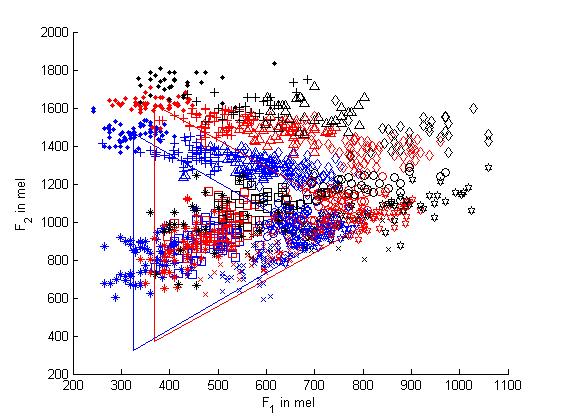}
	\caption{$F_2$ \textit{versus} $F_1$ plot in mel of raw formant data\cite{PnB}. Filled dot:/\textipa{i}/, Plus:/\textipa{I}/, Triangle:/\textipa{E}/, Diamond:/\ae/, Circle:/\textturnv/  Hexagon:/\textipa{A}/ Cross:/\textipa{O}/ Square:/\textipa{U}/ and Star:/\textipa{u}/. Blue: Adult male, Red: Adult female, Black: Children.}
\end{figure}
\begin{figure}
	\centering
	\includegraphics[width=1\textwidth, height=.5\textheight]
	{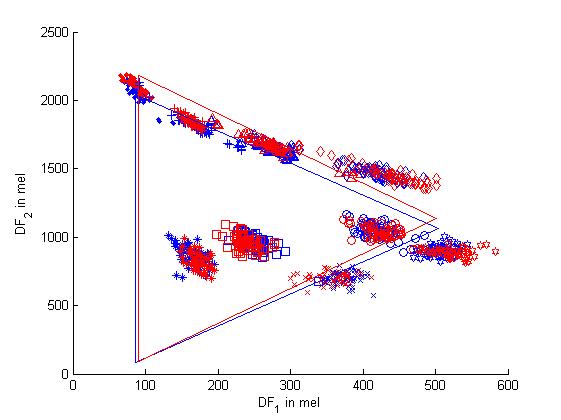}
	\caption{$DF_2$ \textit{versus} $DF_1$ plot in mel of de-normalized formant data obtained using the proposed procedure.}
\end{figure}
\begin{figure}
	\centering
	\includegraphics[width=1\textwidth, height=.5\textheight]
	{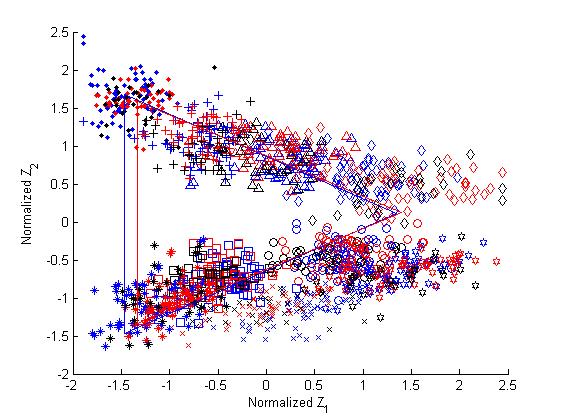}
	\caption{Plot of normalized formant data obtained using the z-score procedure \cite{Lobanov}.}
\end{figure}
\begin{figure}
	\centering
	\includegraphics[width=1\textwidth, height=.5\textheight]
	{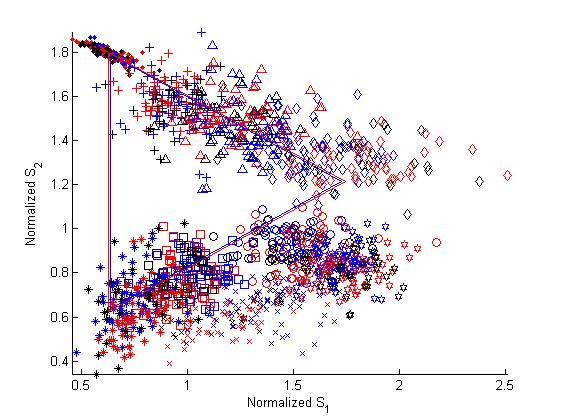}
	\caption{Plot of normalized formant data obtained using the S-centroid procedure \cite{Fabri}.}
\end{figure}
\section{Comparison With Other Methods}
For a comparative purpose we have chosen two top performing \cite{FlynnICPhs, Adank2004} normalization procedures, namely, the z-score \cite{Lobanov} and the S-centroid \cite{Fabri, FlynnICPhs}, \cite{WattFrib}. 
\subsection{Formant Plots and Vowel Triangles}
Plots of normalized formant values using the z-score ($Z_2$ \textit{versus} $Z_1$) and S-centroid ($S_2$ \textit{versus} $S_1$) procedures are shown in Figs. 3 and 4, respectively. The spread in the data points arising due to gender and age difference is reduced for  both the procedures. However, the vowel samples are widely scattered.  In the case of S-centroid procedure, clustering is very good only for vowel /i/ as it acts as a reference corner. It is difficult to infer the number of vowels from the plots shown for z-score and S-centroid. In the de-normalized formant space, one distinct cluster per vowel is seen (Fig.2).

One of the ways to study the effectiveness of a normalization procedure is to compare the overlap of vowel triangles for male (VTM) and female (VTF) speakers \cite{Fabri, FlynnICPhs, WattFrib}. We give only a qualitative comparison. For the raw data (Fig.1), VTF is much bigger than VTM and is significantly displaced upwards and to the right. For the proposed procedure (Fig.2), VTF and VTM almost overlap except for a slight mismatch in the /i/-/\textipa{A}/ direction. For the z-score normalization (Fig.3), VTF is smaller than VTM with a slight mismatch in the /i/-/u/ direction. For the S-centroid method (Fig.4), it is difficult to discern the two vowel triangles as the overlap is almost complete. A vowel triangle is determined by only three normalized parameters, $F_1$, $F_2$ of /i/ and $F_1$ of /\textipa{A}/ and hence it does not reflect the spread of data. We propose to use the accuracy of vowel classification as an objective measure for a comparison of different normalization procedures.

\subsection{Vowel Classification Accuracy as an Objective Measure}
We assume a labeled database of formants of a given language to be available. The set of formant frequencies ($F_1$ $,$ $F_2$) in mel is used as the feature vector. The mean values $\bar\mu_1$ and $\bar\mu_2$ represent the vowel space. Given the test formant data, its nearest vowel in the vowel space is declared as the identity of the test vowel and compared with the known label. The overall accuracy for all the samples is determined. A similar procedure  is applied on the normalized formant values of z-score and S-centroid procedures. Vowel classification is a part of the proposed procedure, as already noted in Sec.II.B. 
We have used a weighted Euclidean distance (WED) measure given by
\begin{equation}
WED^2 = [F_1-\bar\mu_1]^2 / \bar\sigma_1^2 +[F_2-\bar\mu_2]^2 / \bar\sigma_2^2
\end{equation}

\textbf {Selection of test samples:} For the P\&B database, for the gender-independent (MW) case, formant data of \textquoteleft men' and \textquoteleft women' categories and for the gender-age-independent (MWC) case, formant data of all the three categories are pooled together and used. Improvement in vowel classification accuracy, computed with the pooled normalized formant values over the accuracy obtained with the pooled raw formant data is considered as a measure of the effectiveness of the normalization procedure. The vowel dependent statistics ($\bar\mu$, $\bar\sigma$) are computed on the raw and normalized (or de-normalized) pooled formant data using the known labels. For automatic vowel classification, the statistics are to be computed from a training set.

\textbf {Results:} The classification accuracies for the raw data, S-centroid, z-score and the proposed procedures are [82.9\%, 85.0\%, 85.7\%, 95.2\%] for the MW case and [77.2\%, 84.5\%, 84.4\%, 94.9\%] for the MWC case, respectively. The proposed procedure gives the highest accuracy of about 95\%, nearly 10\% higher than the  S-centroid and z-score normalization procedures and 12\% (18\%) higher than the MW (MWC) case of raw data.

\section{CONCLUSION}
We have used vowel dependent statistics and proposed an intrinsic-cum-extrinsic procedure along with a \textquoteleft hypothesize-and-test' paradigm. For the given P\&B database, the large spread observed in the acoustic space for different vowels and talkers has been effectively reduced. Clear clusters have emerged in the de-normalized formant space. The proposed procedure performs better than two top performing procedures in removing the influence of gender and age based on the accuracy of vowel classification as the objective measure. For future work, comparison with other procedures of normalization with rigorous objective measures may be undertaken and the applicabilty of the proposed procedure, over a larger database and in areas like sociolingustics, language change, influence of accent etc. may be explored. The proposed procedure can also be applied on normalized data obtained with other procedures.

\section{APPENDIX}
\subsection{Distortion of formant space due to intrinsic normalization procedure}
Refer to Eq.(2) above.  A plot of $NF_2$ \textit{versus} $NF_1$ in the intrinsically normalized formant space is shown in Fig.5 for P\&B data. The intrinsically normalized back vowels /\textipa{O}/ and /\textipa{u}/ (shown by \lq x' and \lq *', respectively) lie above the vowel /\textipa{A}/ (shown by hexagram) in the /\textipa{A}/ - /i/ direction instead of lying below the vowel /\textipa{A}/ along a line of positive slope. Also, there is considerable overlap amongst the data of different vowels. It is seen that intrinsic normalization results in a distorted vowel space. 

\begin{figure}
	\centering	
	\includegraphics[width=1\textwidth,height=.5\textheight]
	{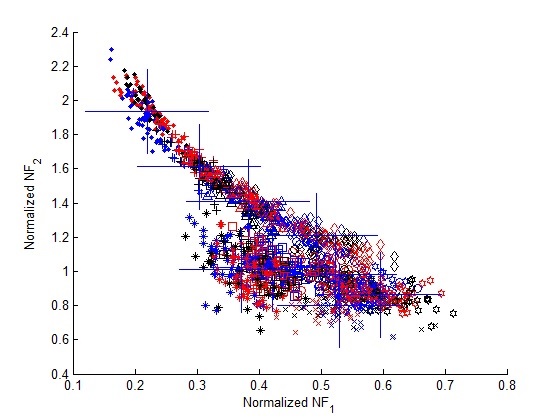}
	\caption{Distribution of formant values, $NF_2$ \textit{versus} $NF_1$ obtained by intrinsic normalization of the Peterson-Barney data. The mean value of each vowel is shown by a cross-hair.}
\end{figure}

\subsection{Restoration of vowel space using the geometric mean of the average values of the first three formant frequencies}
We have defined a vowel dependent factor GMA123 in Eq.(3) as the geometric mean of the \lq \textit{average} values of the first three formant frequencies'. Since the objective of the normalization is to reduce the influence of talker's age and gender, it is to be understood that the \lq \textit{average}' $\mu(i)$ in Eq.(3) refers to the \lq \textit{global average}' of the $i$-th formant frequency obtained by pooling $F(i)$ of all the talkers for the specific vowel. Thus, for each vowel, the ratio  $GMA123$ is used as the factor to obtain the de-normalized $i$-th formant value given by
\begin{equation}	
NNF(i) = NF(i) * GMA123
\end{equation}
The above intrinsic-cum-extrinsic (IE) procedure will be referred to as \lq IE-GMAGM' procedure in contrast to the intrinsic-cum-extrinsic hypothesize-test (IE-HT) procedure. 

A plot of $NNF_2$  \textit{versus} $NNF_1$ is shown in Fig.6 for the P\&B data. This has to be compared with the $F_2$ \textit{versus} $F_1$ plot of raw formant data (Fig.1). There is an improved clustering compared to the raw formant space and the relative locations of the vowels are restored. However, vowels /\textturnv/ and /\textipa{A}/ (represented by circle and hexagram, respectively) overlap heavily. 


To study the effectiveness of clustering obtained using the IE-GMAGM procedure, vowel classification task is carried out on the de-normalized formant values [$NNF_1$, $NNF_2$], assuming the vowel identity as known, as described in Sec.III.B. For the pooled data (MWC), IE-GMAGM procedure gives an accuracy of 91.4\% compared with an accuracy of 77.2\% and 94.9\% for the pooled raw formant data and IE-HT procedure, respectively. The IE-GMAGM procedure gives a significant improvement (14\%) over the raw formant data and only a slightly lower performance (about 4\%) than the IE-HT procedure. The IE-GMAGM procedure is much simpler to implement than the IE-HT procedure. Also, the range of de-normalized formant values matches well with that of the raw formant data.  However, IE-GMAGM procedure is restricted only to acoustic-phonetic studies, where the vowel identity ia assumed to be known and not for automatic vowel classification.


\begin{figure}
	\centering	
	\includegraphics[width=1.0\textwidth,height=.5\textheight]
	{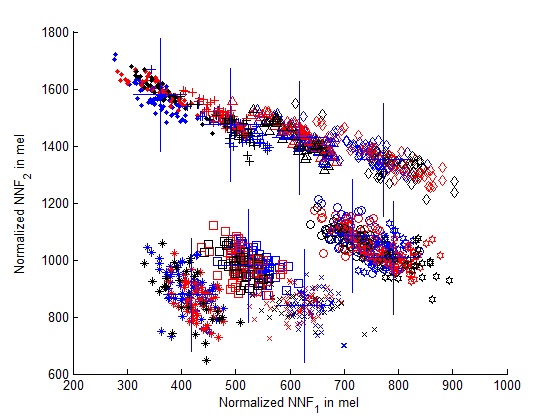}
	\caption{Distribution of de-normalized (IE-GMAGM) formant values (in mel), $NNF_2$ \textit{versus} $NNF_1$ obtained by using the geometric mean of the average values of the first three formant frequencies. The mean value of each vowel is shown by a cross-hair.}
\end{figure}

\subsection{Bootstraping in the \lq Hypothesize-Test' Paradigm}
The purpose of the distance measure of Eq.(5) is to compute the de-normalized formant values and yet the right hand side of Eq.(5) itself uses the mean and standard deviation of de-normalized formant values.  This apparent self-contradiction is resolved by a bootstrapping procedure, which is explained considering the two major application areas of talker normalization, namely (i) acoustic-phonetic studies and (ii) automatic vowel classification. 

In acoustic-phonetic studies, the collected raw formant data, with known vowel identity, is mapped to a new space to study the inter-relationship of vowels after talker normalization. In other words, mapping is a post facto procedure. Hence, for the purpose of acoustic-phonetic studies, the hypothesis \lq $J$' in Eq.(4) is taken directly to be the index of the known vowel. Thus, the mean value of the pooled raw formant data (MWC) of the same vowel $J$ is used for the RHS of Eq.(4), to obtain an initial estimate of the de-normalized formant values of the vowel $J$. Based on these initial estimates, the mean and standard deviation of the \textit {de-normalized} values are computed for each vowel. Subsequently, \lq hypothesize-test' paradigm of Eq.(5) is applied using these statistics to obtain the final estimates of de-normalized values. This bootstrapping procedure has been followed for the illustration in Figure 2 and for vowel classification in Sec.III.B. 

For automatic vowel classification, a labelled training set of samples is assumed to be available.  Initially, the above bootstrapping procedure is followed using the training set data with known vowel identity to obtain the de-normalized values and their statistics. Subsequently, vowel classification is carried out on the test samples of unknown vowel identity and unknown speaker. In the P\&B database, each speaker has uttered each of the vowels twice. We have considered one of these utterances for training and the other for testing. Vowel classification accuracy of about 94.6\% is obtained for the MWC case of test set when statistics of the training set are used. If automatic vowel classification is carried out on the training set using the statistics of the training set itself, an accuracy of 95.8\% is obtained. Thus, there is only a slight (1.2\%) decrease in the accuracy when training set statistics are used for the classification of vowels in the test set. 

The z-score and S-centroid are two top ranking normalization procedures. In both these procedures, the raw formant data of each speaker are translated and scaled so that the vowel space of all the speakers are nearly aligned and of nearly the same size. In z-score normalization, speaker specific mean (for translation) and standard deviation (for scaling) of the raw formant data of all the vowels are used.  There are many different versions of S-centroid procedure. Essentially in S-centroid normalization, speaker specific centroid (for translation) and standard deviation (for scaling) of the three corner vowels are used. In automatic vowel classification, for a given test sample, the speaker identity is unknown whereas procedures such as z-score and S-centroid assume speaker's identity to be known. Hence procedures such as z-score or S-centroid are useful only in acoustic-phonetic studies for mapping the raw formant space and not for automatic vowel classification, even if a training set is available.  

The statistics used in IE-HT procedure are speaker independent and hence the procedure can be applied on a test sample with unknown speaker identity. The vowel and formant dependency of the procedure is overcome by using the hypothesize-test paradigm. IE-HT may be used for both acoustic-phonetic studies and automatic vowel classification (using the training samples).

\subsection{Illustration of Euclidean Distance Computation} 
\subsubsection{Raw formant space}
In Sec.III.B, weighted Euclidean distance measure has been used. However, it is difficult to graphically illustrate the weighted Euclidean distance. Hence, we illustrate only the (un-weighted) Euclidean distance measure. 

In order to appreciate the distance computation used in IE-HT procedure, we initially illustrate the conventional method of Euclidean distance computation in the raw formant space. Figure 7 shows the location of the mean value of each vowel for the pooled raw formant data (MWC) by a cross-hair and the adopted vowel symbol (in red) is shown at the centre of the cross-hair. A randomly chosen raw data sample of vowel /a/ is shown in Fig.6 by a  hexagram (in blue). The line joining the chosen sample and the mean of anyone of the vowels represents the Euclidean distance. In this example, it is seen that the chosen raw data sample is closer to vowel /\textipa{O}/ (shown by \lq x') and hence it is incorrectly assigned.

\begin{figure}
	\centering	
	\includegraphics[width=1.0\textwidth,height=.45\textheight]
	{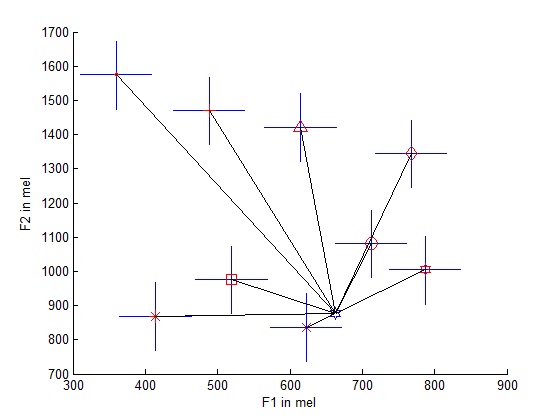}
	\caption{Illustration of the computation of Euclidean distance in the raw formant space. The mean value of each vowel is shown by a cross-hair with the adopted vowel symbol (in red) at the centre. A randomly chosen sample, for the sake of illustration, is also shown (hexagram in blue). The nine lines represent the Euclidean distances between the location of the chosen sample and the mean values of the nine vowels.}
\end{figure}

\subsubsection{De-normalized formant space obtained using \lq hypothesize-test' paradigm}
Figure 8 shows the locations of the mean values of the different vowels for the pooled de-normalized (IE-HT) formant values (MWC) by cross-hairs and the adopted vowel symbols are shown (in red) at the centres of the cross-hairs. Based on Eq.(4), a test sample maps to nine different hypotheses shown (in blue) by the respective vowel symbols. The line joining the location of a hypothesized vowel and the location of the mean of the corresponding vowel represents the Euclidean distance for that hypothesis. There are nine distances corresponding to the nine different hypotheses. For this example, the lowest distance corresponds to the correct vowel identity. 

\begin{figure}
	\centering	
	\includegraphics[width=1.0\textwidth,height=.5\textheight]
	{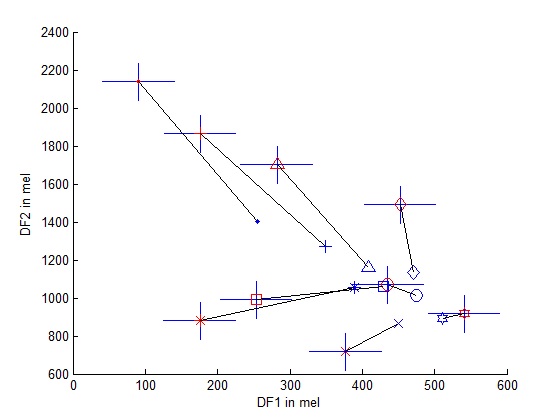}
	\caption{Illustration of distance computation in the de-normalized (IE-HT) formant space obtained using the \lq hypothesize-test' paradigm.  The mean de-normalized value of each vowel is shown by a cross-hair with the adopted vowel symbol (in red) at the centre. The locations of the nine hypotheses (de-normalized by the mean value of the respective formant for the hypothesized vowel) for a chosen sample are shown by the respective vowel symbols (in blue). The nine lines represent the Euclidean distances between the locations of the hypotheses and the mean values of the corresponding vowels.}
\end{figure}

  \end{document}